# A Neural Radiance Field-Based Architecture for Intelligent Multilayered View Synthesis

D. Dhinakaran[1], S. M. Udhaya Sankar[2], G. Elumalai[3], N. Jagadish kumar[4]

[1]Department of Computer Science and Engineering,
Vel Tech Rangarajan Dr. Sagunthala R&D Institute of Science and Technology,
Chennai, India.
*Corresponding Author:drdhinakarand@veltech.edu.in

[2]Department of Computer Science and Engineering(Cyber Security),
R.M.K College of Engineering and Technology,
Chennai, India.
udhaya3@gmail.com

[3]Department of Electronics and Communication Engineering,
Panimalar Engineering College,
Chennai, India.
elumalaig79@gmail.com

[4]Department of Computer Science and Business Systems,
Sri Sairam Engineering College,
Chennai, India.
Jagadishkumar.csbs@sairam.edu.in

**Abstract**—A mobile ad hoc network is made up of a number of wireless portable nodes that spontaneously come together en route for establish a transitory network with no need for any central management. A mobile ad hoc network (MANET) is made up of a sizable and reasonably dense community of mobile nodes that travel across any terrain and rely solely on wireless interfaces for communication, not on any well before centralized management. Furthermore, routing be supposed to offer a method for instantly delivering data across a network between any two nodes. Finding the best packet routing from across infrastructure is the major issue, though. The proposed protocol's major goal is to identify the least-expensive nominal capacity acquisition that assures the transportation of realistic transport that ensures its durability in the event of any node failure. This study suggests the Optimized Route Selection via Red Imported Fire Ants (RIFA) Strategy as a way to improve on-demand source routing systems. Predicting Route Failure and energy Utilization is used to pick the path during the routing phase. Proposed work assess the results of the comparisons based on performance parameters like as energy usage, packet delivery rate (PDR), and end-to-end (E2E) delay. The outcome demonstrates that the proposed strategy is preferable and increases network lifetime while lowering node energy consumption and typical E2E delay under the majority of network performance measures and factors.

**Keywords**-MANET, routing, optimized route selection, network, route failure, energy utilization.

## I. INTRODUCTION

The term "mobile ad hoc network" refers to a wireless connection that spontaneously forms when a number of mobile devices communicate with one another without the aid of an organized system. Since MANETs are spread, nodes must cooperate with one another to ensure the network's operations. However, because nodes in a network are self-organized and also have limited resources, they might act greedily or intentionally to serve their own objectives, for as by refusing to forward packets. Trusting in a node that exhibits poor behavior might result in unanticipated dangers including low network performance, high resource usage, and attack susceptibility [1]. As a result, a trust management is required to let nodes choose how much of many other nodes' activities that may trust. A trust evaluation framework is comprised of elements for knowledge gathering, calculating trust levels, and establishing trust [2-4]. These components work together to create reliable connections throughout the network. Many MANET systems advocate building routes haphazardly by saturating the network with route request (RREQ) packets. Due to the large communication packets caused by the inundation method when creating connections to the target destination, MANET performance may suffer. Additionally, to increase efficiency of the network, flooding processes should indeed be judiciously regulated by lowering the number of mobile nodes transmitting RREQs [5-7]. Rapid network topology reforms introduced on by node mobility can result in frequent link breaks, which add to the network's complexity and create interruptions to the established connections.

259

___

Network performance is strongly impacted by disruption events, which results in greater latency and congestion control and a lower packet delivery ratio (PDR). The need for an efficient link fault prediction approach grows as a result of such problems [8-10]. In order to preserve the paths between both the data source and its matching destination, routing protocols employed in MANET must typically dynamically adjust to different in topology [11-13]. Routing protocols created for mobile ad hoc network are divided into three primary types based on the transit procedure.

(i) The essential characteristic of proactively routing strategies is that each node must frequently communicate route discovery with those other nodes, irrespective as to whether the paths are required.

(ii) The reactive routing procedures; in these, an origin merely searches for a path to an endpoint when it is necessary.

(iii) Hybrid routing protocols, which combine the positive aspects of reactive and proactive procedures. In a dynamic world with common topological changes, the hybrid and proactive protocols won't provide acceptable results in terms of operating costs and memory reducing due to their sluggish diagnosis of and reacts to path breaks and the superfluous transact of constant updating.

To conserve bandwidth across the network, on-demand schemes were created to reduce the amount of control messages sent. A route to an endpoint is only looked up when the subsequent protocol layers demand it. The two types of reactive routing protocols are hop-by-hop and source-based routing [14-17]. Source-based routing systems like dynamic packet forwarding (DSR) hold the whole route to the target, in contrast to hop-by-hop routing protocols, which only carry the target and the next hop addresses in respective packet data headers [18]. DSR, asserts itself as a legitimate direction-finding system with such qualities, is one of most widely acknowledged on-demand routing algorithms. Routing protocol and routing information are the two primary methods used by DSR; both of these mechanisms work when a path is in demand. Alternative courses are, however, typically found by inundating the network with route request packets that move endlessly over the whole network [19-22]. Therefore, it is important to control inundating activities only when they are effective and valuable for the network. Additionally, the effectiveness of the network is affected by the regular link failures caused by node occurrences, which raises the need for an efficient failover prediction technique.

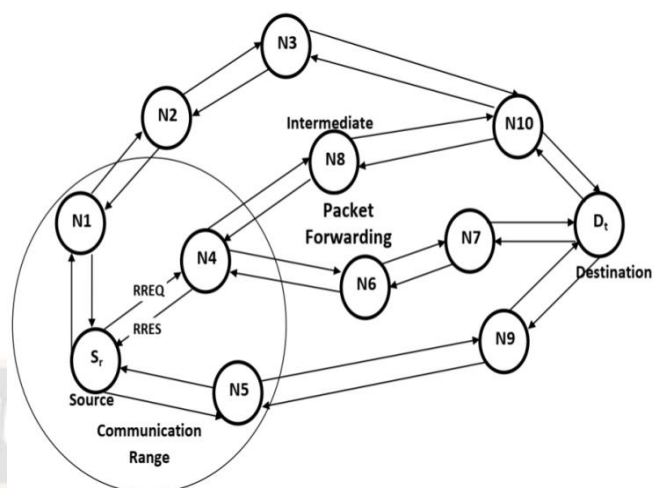

Figure 1: Scenario of Classic MANET System

N mobile nodes in a MANET that communicates utilizing vertex-based wireless edges (V). A graph G D is created when the nodes are connected in a random-mesh topology (N; V). The availability of V is determined by the interaction range R and the distance d here between nodes [23]. In Figure 1, a typical MANET situation is shown. The neighbor node that makes up the different routing path are used by the origin to send packets to the destination [24 -26]. Since the origin and destination are separated by a great distance, multi-hop transmission is used. We presumptively move the nodes in a randomly, which has less of an impact on how the network works. The assault mode is not described in this circumstance because that article focuses on trustworthy buddy choosing. The proposed study leverages the unique Optimized Route Selection Via Red Imported Fire Ants (RIFA) Approach to improve on-demand resource routing protocols. Predicting Route Failure and Power Utilization is used to pick the path during the routing phase.

## II. LITERATURE SURVEY

Since they are made up of mobile nodes that communicate across wireless links without centralized supervision, ad hoc networks are vital to the development of wireless mobile networks [27-30]. Ad-hoc wireless networks immediately inherit the issues with standard wireless and mobile networking, such as maximizing bandwidth, improving transmission quality, and controlling power [31]. Due to their multi-hop structure, presence of a fixed infrastructure, ad hoc routing, and self-routing, ad hoc networks are also responsible for new research issues including Configuration adverts, detection, and preservation [32-35]. In the Internet Protocol, there are various standardizing initiatives underway as well as numerous commercial and academic endeavors, in addition to a lot of proposals on various techniques and protocols.

___

For mobile edge computing, D Zhang et al. [7] offer LLECP-AOMDV, an ad hoc on-demand multi-path distance vector (AOMDV) routing protocol based on connection duration and prediction of energy consumption. Whenever the energy of the node's cascade underneath a fixed threshold, it discontinues taking part in routing protocol. Choosing the path in the phase of routing selection is based on the life span of the network link and the path's lowest energy usage. Kacem et al. [8] concentrate on figuring out the best traffic flow over the system. The proposed protocol's major goal is to identify the least-expensive nominal capacity expenditure that guarantees the channeling of hypothetical traffic and assures its durability in the event of anynode malfunction. Throughout this aspect, the ant system is employed to discover a way to address the crisis of ambiguity measures in MANET, and the Petri net is largely utilized for navigation and recognition operations via synchronous fuzzy transitioning strategy.

The secure neighborhood selection method employing recurrent incentive learning is highlighted by Sankaran et al. [9] proposed a method for identifying node stages based upon communication activity incorporates the advantages of traditional routing and clever machine learning paradigms. Establishing stable and safe routing and transmitting channels to the destination requires in-depth knowledge of the nodes' performance at all connection hop-levels. A novel Obstacle-aware and Manoeuvrability Routing (OMAR) strategy was proposed in paper [10]. This plan uses the DeCasteljau approach with a Bezier curve to detect collisions. A Power reliant Migration Index forwarding method utilized to reduce the effects of mobility of a mobile node. The route to the destination is selected based on which has the highest EMI value. But they have examined a few rectangular obstructions.

In each of these categories, Rajeswari et al. [11] present a survey to review comparative research of multi - hop routing methods. There are also brief analyses of significant routing problems. This review study evaluates the characteristics of communication algorithms and emphasizes on the taxonomy associated with MANET. Ochola et al. [12] focuses on the black hole assault on MANET reacting routing protocols has been examined in this research (AODV and DSR). To analyze the impact of node may be determined that AODV is preferable in a mobility network because simulations findings forecast the effectiveness of DSR declines more quickly than the effectiveness of AODV whenever the velocity of the terminals is enhanced.

Relay routed DSR is an improvement put out by Shobha and Rajanikanth [13] to cut down on the number of RREQ and control packets. This approach chooses the intermediate nodes to which RREQs should indeed be transmitted during the mediating phase based on the transportation information gathered from the nearby nodes during flood process. However, due to the velocity of the migrating node, leading to repeated route explorations that add additional overhead. In paper [14], researchers suggested a DSR-based scheme in which the Continuous Hopfield Neural Network (CHNN) is used to determine the network link's sturdiness to obtain the path with the greatest stabilization of the carefully crafted the data transmission end to the receiver side node of the DSR protocol. The results of the test show that CHNN-DSR performs better than DSR in contrast. However, they did not incorporate actual terrain characteristics into their simulations.

A unique context-aware search navigation technique for unorganized P2P wireless file transfer systems is presented by Sofian et al. [15] their protocol finds relevant peers who are sharing resources that are relevant to the client query; and (ii) makes sure that clients would be accessible by taking distinct MANET restrictions into account. They build our approach on the technique for ordering desires by resemblance to optimal situation in order to take into account these all limitations when selecting the pertinent peers. A novel routing strategy called the energy awareness fuzzy control router (EAFCR) algorithm is proposed by Helen et al. [16] During the exploration process, the proposed algorithm applies fuzzy tools to create a secure and energy-efficient path, enhancing the cognition of the nodes. The fuzzy system creates a much more reliable routing by using the per hop latency, energy availability, and signal strength.

With the goal of enhancing the routing protocol effectiveness in MANETs, Chalew et al. [17] created and refined the manoeuvrability routing algorithm (MARA). The suggested system enables access points to rebroadcast or ignore messages that have already been broadcast. The choice is made based on the interaction among node speed, node length, and junction based on residual. To lessen the possibility of link failure and broadcasting storm issues, these variables are taken into account during the routing path and route reply processes. Using the intensity of the acknowledged packets and the preemptive DSR (PDSR) protocol, Ramesh et al. [18] predicted the link breakage time. In this method, the source node creates two routes—the main route as well as the backup route—during the process of routing. The intermediate nodes on the alternate path continue monitoring the signal received throughout the conversation period. The intermediary node transmits a clear warning to the intermediate host if this emission strength falls underneath the threshold level.

The alternative route is used by the source and destination when it first obtains the clear warning; whether this channel fails as well, the origin node forwards a fresh route discovery process. Unfortunately, since it has to conduct route changing more regularly whenever the network topology shifts frequently, PDSR reacts slowly. To deal with node failure, PDSR uses a slow and expensive link fault prediction

**261**

___

technique. In Paper [19], energy back-off network variable is combined with a new Energy- docile Router strategy to improve it. The proposed approach improves the node-life duration and the effectiveness of MANET under different traffic conditions and power levels by combining with different power generating methods. Nevertheless, they did not incorporate actual terrain characteristics into their simulation.

### III. SYSTEM MODEL

*A. Motivation*

Ad hoc network interactions are mostly dependent on radio signal. Despite the fact that the wireless ecosystem is quite flexible, it has many drawbacks, including frequent service outages and changeable debits. Although new methods have been created to solve these issues, some remain exist in spite of cost-cutting and downsizing efforts. These issues mostly revolve around endpoint mobility, where examine the effect can result in the loss of connectivity.

The works that have already been done use optimization techniques. The suggested technique uses the Red Imported Fire Ants (RIFA) Approach to accomplish Optimized Route Selection. Predicting Route Failure and Power Utilization is used to pick the path during the routing phase. Considering bandwidth limitations, longevity, and reliability, we want to increase the Packet Delivery Ratio (PDR), increase network capacity, and decrease end-to-end (E2E) latencies.

The following list of facts summarizes this article's key contribution:

- Using the RIFA Approach, defining the different types of nodes that were present in the network at period t as well as the alterations that each node underwent.

- Setting forth a number of parameters or signs (e.g. PDR, network capacity, higher energy node, delay, and distance).

- Depending on the RIFA, selecting the optimal nearby node depending on the trust value offered by nearby nodes.

This process, which suggests a new detection method based on the termite's mechanism, seeks to lower the additional cost of communicating.

- The route discovery protocol's inclusion of the detection technique enabled the reduction of interrupted paths and minimize reciprocal inferences among nodes. The Packet Delivery Ratio and the network lifespan (LT) are raised, and the E2E latencies are decreased, by using a diversity of network indicators.

The RIFA pheromone is incorporated into MANET operations as part of the suggested strategy. The least range nodes are found through ad-hoc routing protocols to create routes to the target. The route construction and neighborhood choice in this pheromone are aided by other node properties other than distance. By including effective neighbors, the multi-attribute assessment boosts the validity of the chosen path. As a result, the suggested approach is broken down into three phases: route selection, energy consumption prediction, and link failure prediction.

*B. Red Imported Fire Ants Approach - Route Selection*

The severe sting of red imported fire ants (RIFA), which are more antagonistic than other ant species. The ability of red imported fire ants to adapt to various climatic and environmental circumstances. Ants are tackling a provides a potential and optimizing problem that is still largely unexplored when they engage in cooperative hunting in ant colonies, which is equally astonishing. As they move, ants leave behind pheromone, and subsequent ants prefer to follow trails that have left behind more pheromone [36]. Ants swiftly give up on other pathways to focus on the lowest one without ever speaking to one another directly. If the food source shifts or a new road is found, ants can still migrate to the different path despite initial bias caused by pheromones left on an earlier trail. When there are several food sources available, ants create numerous routes that appear to improve total throughput, if such a term is appropriate.

\_\_\_\_\_\_\_\_\_\_\_\_\_\_\_\_\_\_\_\_\_\_\_\_\_\_\_\_\_\_\_\_\_\_\_\_\_\_\_\_\_\_\_\_\_\_\_\_\_\_\_\_\_\_\_\_\_\_\_\_\_\_\_\_\_\_\_\_\_\_\_\_\_\_\_\_\_\_\_\_\_\_\_\_\_\_\_\_\_\_\_\_\_\_\_\_\_\_\_\_\_\_\_\_\_\_\_\_\_\_\_\_

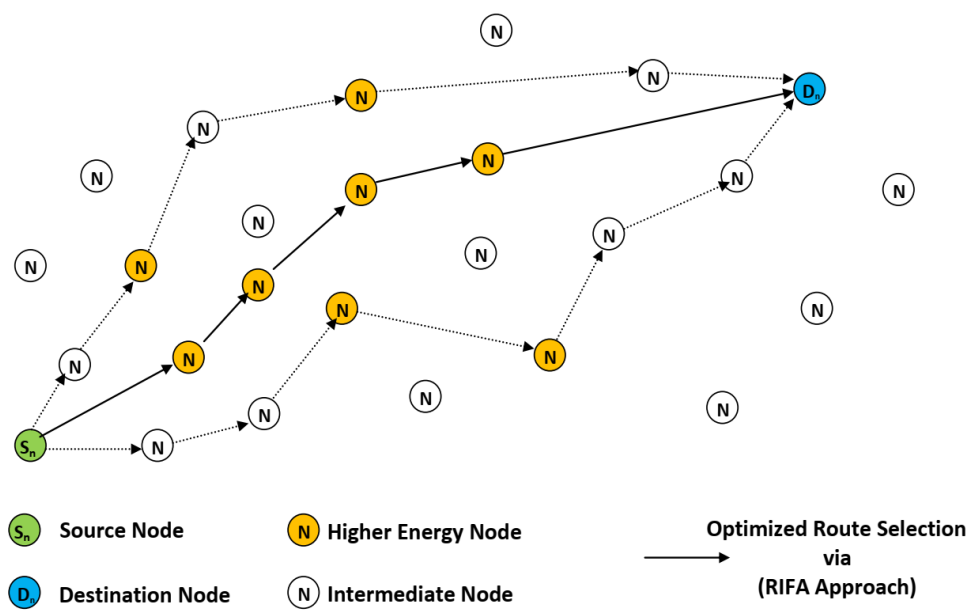

Figure 2: Architecture of Proposed Model

One of the most amazing examples of cooperative processing is seen in the communal scouting of ants. Ants leave behind pheromone when moving, and they only use local knowledge and inferential interactions to spread their judgments. Ants favor a route with a superior pheromone awareness when there are several other options available. A route with a superior pheromone attentiveness likely to be favored by subsequent ants and one that was used more frequently by earlier ants. Given that the optimization is accomplished without explicit knowledge of the place with high of one route over the other, an unified command-and-control framework, or explicit ant-to-ant interaction, this phenomena is extraordinary. In recent years, optimizing the ant colony has become a significant dispersed optimization algorithm [37].

Table 1: Notations and Meanings

| S. No | Notation | Meaning |
|---|---|---|
| 1 | n | Independent Alternate Paths |
| 2 | t | Units Of Pheromone - Path |
| 3 | Cd | $K^{th}$ Path |
| 4 | a | Distribution |
| 5 | b | Time Instant |
| 6 | $\Gamma(\cdot)$ | Gamma Function |
| 7 | d | Path |
| 8 | C | Starting State |
| 9 | p | Sender Node |
| 10 | q | Next-Hop Node |
| 11 | (xp, yp) | Location Coordinates |
| 12 | (vp$\theta$p) | Speed |
| 13 | Tr | Transmission Range |
| 14 | Rd | Remaining Distance |
| 15 | Ti | Time Interval |

*C. Shortest Route*

Based on the presumption that there are n distinct alternative routes between the food supply and the nest. by presuming that ants select a path based on probabilities from those that are accessible. Additionally, an ant selects a path and only releases t components of pheromone along that path, never releasing any along any other paths. Since only one ant may make a decision at a time, it is supposed that ants make decisions asynchronously as shown in Fig. 1. It is believed that ants only release pheromone on their way back. In the follow-up, this presumption is revisited. By presuming that when an ant chooses a path, it follows it until it reaches its destination, using it to get to the source of food and afterwards return to a nest [38-40]. The set of n possible paths is comparable to the vase filled with n distinct coloured balls. The quantity of balls of colour j in the vase correlates to the scent intensity on the $j^{th}$ route. An ant randomly selects one of the routes at each stage, which is analogous to selecting a ball from of the vase. The number t updates the pheromone intensity along the selected path, which is comparable to putting t balls of colour j to the vase.

After some time, the returning path's pheromones update. Ants traveling at the same speed will face the very same latencies if all routes have the same size. This scenario, which is a perfect analogue to the Polya Urn Process, is what we look at first. They will face various delays whenever the path lengths are variable. The reader might turn to for more information because we do not cover the latter scenario here. Consider the scenario where n possible paths exist. This is comparable to the situation where the urn contains several distinct coloured balls.

**263**

___

Cd, 1 I k stands for the initial scent intensity on the k-th route. The distribution "a" at time instantaneous b, originating from the initial state C, is given by equation (1).

$$P(a,b,c) = \frac{\Gamma(\sum_{d=1}^{n} c_d/t)}{\Gamma(b+\sum_{d=1}^{n} c_d/t)} \prod_{d=1}^{n} \frac{\Gamma(a_d+c_d/t)}{\Gamma(c_d/t)} \quad (1)$$

where the Gamma function is denoted by $\Gamma(\cdot)$. An approximate solution which can be stated through using Stirling's equation after adding up all the possibilities that route d is selected xd instances, providing that b as well as an are substantial. A mobile node could transmit one of two statuses to its neighbors:

$$\Gamma(b) = \sqrt{2\pi(b-1)} * \left(\frac{b-1}{x}\right)^{b-1} \quad (2)$$

The common rule for $a = \{a1, a2, \ldots, ak\}$ updates on paths $\{1, 2, \ldots, n\}$, with initial distribution $C = C1, C2, \ldots, Cn$ is given by

$$P(a,b,c) = \frac{\Gamma(\sum_{d=1}^{n} c_d/t)}{\prod_{d=1}^{n} \Gamma(c_d/t)} \cdot \frac{\prod_{d=1}^{n} a_d^{\frac{c_d}{t}-1}}{\sum_{d=1}^{n} \frac{c_d}{t}-1} \quad (3)$$

The well-known Beta distribution serves as the probability density function in the two-path example. The starting pheromone intensities on the two pathways match the Beta distribution's form characteristics. The Dirichlet distribution, that is an extension of the Beta Distribution, is what we get in the k-path scenario, though. These findings demonstrate that the baseline pheromone intensity alone determines the likelihood of trail creation on a path.

*D. Energy Consumption Prediction*

In our protocol, the energy needed for data transmission and the link's lifespan are factored in during the path selection stage. Routing Request (RREQ), Routing Reply (RREP), and Route Error are the three input packets in the protocol (RERR). Overlooked are the node's remaining energy reserves and the path's data utilization. Neglecting the path's reliability, it is feasible to construct the chosen way with the fewest hops, but the length or energy usage are great, increasing the likelihood of link separation. Therefore, in this section, we suggest a prioritized path selection algorithm [41-43]. When path s is chosen, the network's lifespan as well as the path's consistency are improved by carefully taking into account the route's life time experience and energy consumption during transmission.

**Algorithm - Mobile Edge Computing**

1) To send data - source,
   Check routing information available
   If (there are no route details involved)
   {       start - routing request
           send - Routing Request - (RREQ)
   }
2) If an intermediate node receives the RREQ,
   Judges its own energy level
   Calculate the R
   Judges the level of its own energy.
3) While (node - Danger level)
   Directly discarding RREQ;
4) While (node - Normal level)
   Node determines if a viable alternative reversing path exists
   Route will be update
   Check if
   {       forwarding route to the target node
           In none of the earlier RREPs, the forward route is used
   }
   If (present node generates - RREP)
   The intermediate node does not continue to disseminate RREQ
   Else If (no direct route to the target node)
   Periodically transmits the initial RREQ copy it has obtained
5) Destination will generate a RREP IF
   {
   RREQ copy is received by a destination
   Reverse path is formed by the destination node
   RREQ copy comes from a different neighbor
   }
6) To generates a RREP,
   Determine if the RREP is being sent on a backward pass in this multipath routing
   If (exist)
   {
   To submit the prevailing RREP, pick one of the unused alternatives
   }
   else
   {
   castoffs the RREP
   }
7) If (RREP received by the source)
   Create the routing table
   Calculate R
8) The obtained multiple paths are link-disjoint and loop-free
9) While transmitting data, a route with a top importance is used.

*E. Link Failure Prediction*

Link failure prediction makes use of mobility data, node density, the range to travel before leaving the originator node's scope region, and Hello message interval. The transmitter shall frequently check the implements until the next hop for each

___

Hello periods for two interacting participating nodes in the alternate path from the source host to the destination [44]. The suggested link fault prediction strategy makes use of specified threshold settings and mobile data, with each node's mobility data being received via GPS. Whenever the link towards the nexthop nodes is due to fail, the suggested Failover detection is activated.

In order to start the process of finding a new route to the desired destination before connection failure happens, the transmitter node will send an acknowledgment message to the origin node. The dynamic path from of the origin point in the direction of the target network consists of two connected nodes travelling at randomized routes and speeds [45]. The sender node, designated as node p, has the positions (xp, yp), and its pace is (vpθp). The next-hop node, Endpoint q, has the positions (xq, yq), and its pace is (vqθq). Rd is the residual range before node q becomes node p's next-hop. Tr is the broadcast range. The variables utilized in the creation of a statistical model of link defect prediction for pair of nodes, p and q, are explained in the following subcategories.

*F. Hello Message Interval*

In order to keep each other informed about node condition, position coordination, and transportation info, all nodes communicate a Hello packet among respective neighboring node. Every node of the network that uses zone-based route to the destination keeps a neighboring table that contains all information about its neighbours and divides them into three zones based on where they are in relation to the node's communication range. By transmitting a Hello packet at predetermined intervals. It is crucial to include this variable in the link failure forecasting model because the sender node can get latest data about its neighbours throughout the Hello message time interval (Ti).

*G. Link Stability*

Link stability, which determines how long a linking between two linked nodes can remain stable, is calculated in part by taking into account link expiry time. Due to their importance in establishing link expiry time, link reliability and link expiration time can therefore be regarded as the structure's key terms. The following formula is used to determine the link expiry period between two nodes, I and j:

$$\text{link expiry time} = \frac{-(pq+rs)+\sqrt{(p^2+r^2)tr^2-(ps-qr)^2}}{p^2+r^2} \quad (4)$$

### IV. PERFORMANCE ANALYSIS

To assess the effectiveness of the proposed mechanisms, we conducted experiments using a simulation platform, as detailed in Table 2. We created a random sample of source nodes, with node counts ranging from 20 to 125. These source nodes were configured to move at varying speeds, ranging from 20 m/s to a maximum of 25 m/s, with intermittent stop periods that ranged from 0 to 50 seconds. These nodes transmitted data packets at a constant bit rate (CBR). The transmission range for these nodes was set at 200 seconds. We employed the random waypoint (RWP) technique to generate the movement patterns of different nodes. We ran simulations for a duration of 600 seconds, which was deemed sufficient to evaluate network congestion, latency, and complexity. For each alternative route, we computed the "Optimal Path Parameter" (OPI) based on various factors such as Movement Indication, Path Accessibility, Link Duration, and Network Accessibility. When transferring data from intermediate nodes to the destination node in the Mobile Ad Hoc Network (MANET), we selected the route with the highest Path Selection Factor. This approach allowed us to thoroughly analyze and assess the performance of our proposed mechanisms.

Table 2: Simulation Parameters

| Parameters | Values |
|---|---|
| Number of Nodes | 20 - 125 |
| Simulation Time | 600s |
| Packet size | 512 bytes |
| Speed of the node | [20-25]m/s |
| Mobility Model | Random |
| Wireless Range of Transmission | 200s |
| Protocol used for Routing | RIFA |
| Traffic category | CBR |
| Area of Simulation | 1200m |
| Pause Time (s) | 10s |
| Node Assignment | Random |

*A. Packet Delivery Rate*

The packet delivery rate (PDR) is a measure of performance defined as the ratio between the number of packets generated by higher layers and the number of packets successfully received at their intended destination. This standard serves as an indicator of the effectiveness of the proposed method for transmitting data from a source to a destination. It's worth noting that the proposed method's performance improves as the rate of successful data packet delivery increases [34,35]. For the sake of simplicity, we'll represent the data packet delivery performance as PDR, which is determined using the following formula:

$$\text{Packet Delivery Rate} = \frac{N_{pr}}{N_{ps}} * 100\% \quad (5)$$

where $N_{pr}$ stands for the quantity of packets received while $N_{ps}$ for the quantity of packets transmitted.

_______________________________________________________________________________________________________________________________

| Table 3: Variation of Packet Delivery Ratio | | | | |
|---|---|---|---|---|
| **Nodes** | **12[RRP]** | **14 [CHNN]** | **17[MAR]** | **Proposed Approach** |
| **20** | 0.88 | 0.91 | 0.92 | 0.95 |
| **40** | 0.85 | 0.89 | 0.88 | 0.91 |
| **60** | 0.83 | 0.86 | 0.85 | 0.89 |
| **80** | 0.80 | 0.84 | 0.82 | 0.85 |
| **100** | 0.78 | 0.81 | 0.80 | 0.84 |
| **120** | 0.75 | 0.79 | 0.77 | 0.82 |

In Figure 3, we observe the fluctuation in packet delivery ratios (PDR) for different methods/protocols: Reactive Routing Protocol (RRP), Continuous Hopfield Neural Network (CHNN), and Mobility-Aware Routing (MAR). As the speed of nodes increases, the PDR decreases. Specifically, for the proposed strategy, the PDR drops from 95% to 82%, for RRP, it decreases from 88% to 75%, and for MAR, it declines from 92% to 77%. Notably, compared to the previous protocols, the proposed method exhibits a higher PDR, indicating a more reliable approach. The suggested routing protocol prioritizes selecting the most dependable route to the destination. This chosen path may possess the highest energy level, consume the least amount of energy compared to alternative routes, and be the shortest in terms of distance. Consequently, this approach lowers the likelihood of node failures and minimizes data loss.

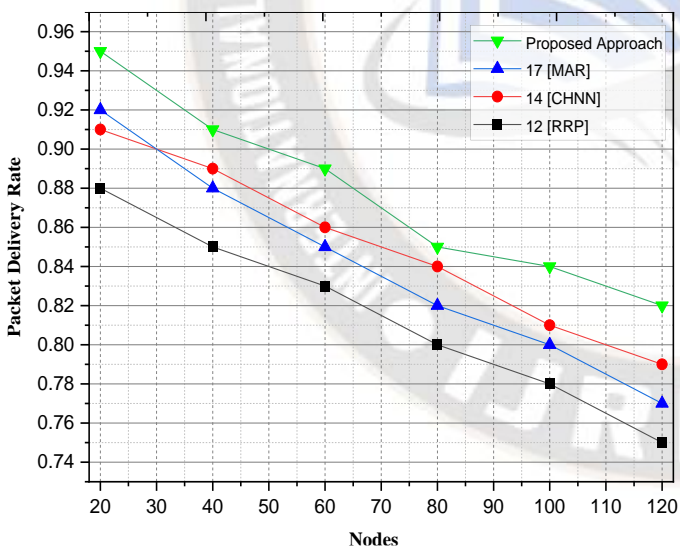

Figure 3: Variation of Packet Delivery Ratio

### B. Average End-To-End Delay

The averaged end-to-end (E2E) latency represents the duration it takes for a data packet to travel effectively from one point to another [39-42]. To denote this typical end-to-end delay, we use the abbreviation E2ED. The computation process is described as

$$E2ED = \frac{1}{Tdp}\sum_{a=1}^{Tdp}(T_{pt}(a) - T_{pr}(a)) \qquad (6)$$

where Tdp stands for the total number of requires very high data packets, $T_{pt}(a)$ for the a[th] packet's transmission time, and $T_{pr}(a)$ for the packet's reception time.

Table 4: Average End-To-End Delay of the Proposed Approach

| Node pause time (Sec) | 12[RRP] | 14 [CHNN] | 17[MAR] | Proposed Approach |
|---|---|---|---|---|
| **15** | 9 | 9 | 8 | 7.5 |
| **25** | 13 | 14 | 12 | 10 |
| **35** | 16 | 19 | 16 | 14 |
| **45** | 21 | 22 | 19 | 17 |
| **55** | 24 | 26 | 23 | 19 |
| **65** | 28 | 29 | 27 | 22 |

In Figure 4, we can observe a comparison between the average end-to-end (E2E) delay of the suggested approach and that of the Reactive Routing Protocol (RRP), Continuous Hopfield Neural Network (CHNN), and Mobility-Aware Routing (MAR). Notably, as the time duration increased, the end-to-end latency decreased. Consequently, the suggested method exhibited a notable latency of approximately 17 ms when the pause period was 45 seconds. However, as the pause duration increased further, the E2E delay decreased, attributed to reduced transit time and a lower likelihood of node failures.

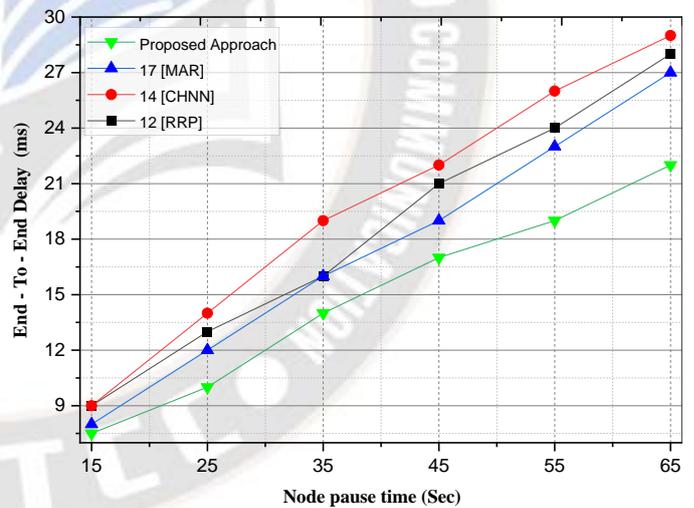

Figure 4: Average End-To-End Delay of the Proposed Approach

### C. Energy Consumption

Energy consumption in the network is defined as the total amount of energy expended by network nodes throughout the scenario. To calculate this, we assess each node's energy level at the end of the experiment, taking into consideration its remaining energy. In this context, we will represent energy consumption using the term "Econs," which can be computed as

$$Econs = \sum_{a=1}^{B}(Ei(a) - Er(a)) \qquad (7)$$

**266**

___________________________________________________________________________________________________________________

Table 5: Energy Consumption - Node Speed

| File Size (Kb) | 12[RRP] | 14 [CHNN] | 17[MAR] | Proposed Approach |
|---|---|---|---|---|
| 2 | 83 | 78 | 69 | 63 |
| 4 | 89 | 82 | 73 | 70 |
| 6 | 102 | 98 | 83 | 79 |
| 8 | 112 | 107 | 90 | 84 |
| 10 | 121 | 113 | 98 | 88 |
| 12 | 128 | 117 | 103 | 91 |

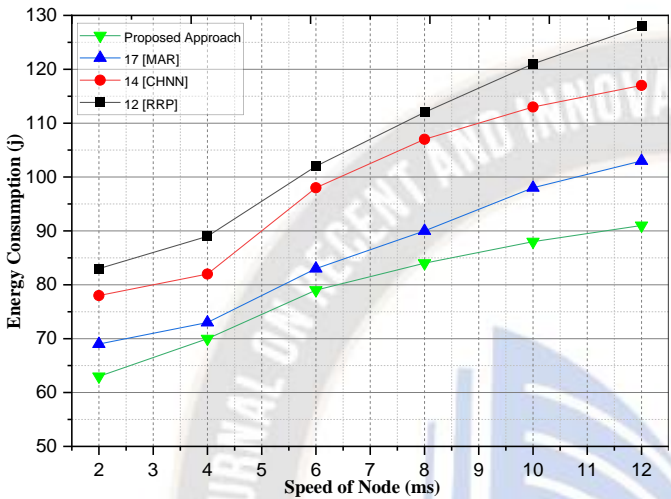

Figure 5: Energy Consumption - Node Speed

Table 6: Energy Consumption - Simulation Time

| File Size (Kb) | 12[RRP] | 14 [CHNN] | 17[MAR] | Proposed Approach |
|---|---|---|---|---|
| 10 | 31 | 31 | 32 | 28 |
| 20 | 60 | 66 | 70 | 50 |
| 30 | 66 | 71 | 82 | 60 |
| 40 | 80 | 85 | 94 | 70 |
| 50 | 90 | 94 | 104 | 83 |
| 60 | 95 | 105 | 121 | 91 |

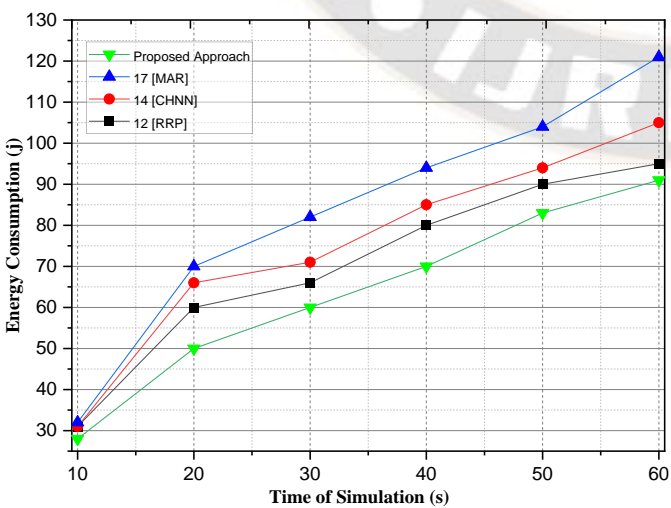

Figure 6: Energy Consumption - Simulation Time

In our study, we have proposed three different approaches: Reactive Routing Protocol (RRP), Continuous Hopfield Neural Network (CHNN), and Mobility-Aware Routing (MAR), and we've examined the variation in their energy usage, as illustrated in Figure 5. It's worth noting that energy usage increases as the speed of nodes increases. Specifically, in the case of the proposed method, energy usage rises from 63 to 91 joules, RRP sees an increase from 83 to 128 joules, CHNN exhibits a rise from 78 to 117 joules, and MAR experiences an increase from 69 to 103 joules. In comparison to the alternative protocols, our suggested protocol demonstrates more efficient energy utilization.

The proposed technique classifies the paths to the target based on their energy levels. To evenly distribute the traffic load across multiple routes, the source node disperses data packets through paths with both high and average energy levels. This approach results in lower energy consumption compared to routing all the traffic through a single route. In Figure 6, we present the energy usage of the recommended approaches, which include Reactive Routing Protocol (RRP), Continuous Hopfield Neural Network (CHNN), and Mobility-Aware Navigation (MAR). Notably, the suggested technique consumes 55 joules over 20 seconds and 83 joules over 50 seconds, while other protocols consume more than 60 joules over 20 seconds and 85 joules over 50 seconds. It is evident that the proposed approach utilizes energy more efficiently than some other strategies.

## V. CONCLUSION

A mobile ad hoc network (MANET) is made up of a sizable and reasonably dense community of mobile nodes that travel across any terrain and rely solely on wireless connections for communication, not on any centralized management. The key challenge with MANET is figuring out the best way to route packets throughout the network. This study suggests a brand-new routing mechanism for ad hoc networks. The fundamental concept is to examine the confidence levels for each communication link in order to provide an adaptable strategy to lessen congestion and end-to-end latency.

The suggested protocol's method for determining the optimum paths is centered on the RIFA protocol. Predicting Route Failure and Power Utilization are used to pick the path during the routing phase. The following benefits come with this suggested approach to controlling the flow in network and the proposed technique be able to identify link failures and swiftly suggest new forwarding table, to keep away from significant transmission delays that result in loss of packet. Utilizing performance metrics like energy consumption, packet delivery rate (PDR), as well as end-to-end (E2E) delay, we review the effectiveness of the assessments. The outcome demonstrates that the proposed approach is better under the majority of

___

network performance characteristics and indices, increases lifetime of the network, decreases node energy usage, and decreases average end-to-end delay.

## REFERENCES

[1] G. Liu, Z. Yan, and W. Pedrycz, Data collection for attack detection and security measurement in mobile ad hoc networks: A survey, J. Netw. Comput. Appl., 105 (2018) 105–122.

[2] D.-G. Zhang, J.-N. Qiu, T. Zhang, and H. Wu, 'New energy-efficient hierarchical clustering approach based on neighbor rotation for edge computing of IOT, in Proc. ICCCN, Valencia, Spain, 8(1) (2019) 291–295.

[3] Dhinakaran D, Joe Prathap P. M, "Protection of data privacy from vulnerability using two-fish technique with Apriori algorithm in data mining," The Journal of Supercomputing, 78(16), 17559–17593 (2022).

[4] S. Glass, I. Mahgoub, and M. Rathod, Leveraging MANET-based cooperative cache discovery techniques in VANETs: A survey and analysis, IEEE Commun. Surveys Tuts., 19(4) (2017) 2640–2661.

[5] D. Dhinakaran and P. M. Joe Prathap, "Preserving data confidentiality in association rule mining using data share allocator algorithm," Intelligent Automation & Soft Computing, vol. 33, no.3, pp. 1877–1892, 2022.

[6] L. Srinivasan, D. Selvaraj, S. M. Udhaya Sankar, "Leveraging Semi-Supervised Graph Learning for Enhanced Diabetic Retinopathy Detection," SSRG International Journal of Electronics and Communication Engineering, vol. 10, no. 8, pp. 9-21, 2023.

[7] D. -G. Zhang et al., A Multi-Path Routing Protocol Based on Link Lifetime and Energy Consumption Prediction for Mobile Edge Computing, in IEEE Access, 8 (2020) 69058-69071, doi: 10.1109/ACCESS.2020.2986078.

[8] Kacem, B. Sait, S. Mekhilef and N. Sabeur, A New Routing Approach for Mobile Ad Hoc Systems Based on Fuzzy Petri Nets and Ant System, in IEEE Access, 6 (2018) 65705-65720, doi: 10.1109/ACCESS.2018.2878145.

[9] S. Sankaran, N. Vasudevan, K. R. Devabalaji, T. S. Babu, H. H. Alhelou and T. Yuvaraj, A Recurrent Reward Based Learning Technique for Secure Neighbor Selection in Mobile AD-HOC Networks, in IEEE Access, 9 (2021) 21735-21745, doi: 10.1109/ACCESS.2021.3055422.

[10] B. K. Panda, U. Bhanja, and P. K. Pattnaik, Obstacle and mobility aware optimal routing for MANET, J. Intell. Fuzzy Syst., 37(1) (2019) 1007–1017.

[11] R. Rajeswari, A Mobile Ad Hoc Network Routing Protocols: A Comparative Study, in Recent Trends in Communication Networks. London, United Kingdom: IntechOpen, (2020). doi: 10.5772/intechopen.92550.

[12] E. O. Ochola, L. F. Mejaele, M. M. Eloff and J. A. van der Poll, Manet Reactive Routing Protocols Node Mobility Variation Effect in Analysing the Impact of Black Hole Attack, in SAIEE Africa Research Journal, 108 (2) (2017) 80-92, doi: 10.23919/SAIEE.2017.8531629.

[13] Shobha. and K. Rajanikanth, Efficient flooding using relay routing in on-demand Routing protocol for Mobile Adhoc Networks, in Proc. IEEE 9th Malaysia Int. Conf. Commun.(MICC), (2009) 316–321.

[14] H. Yang, Z. Li, and Z. Liu, A method of routing optimization using CHNN in MANET, J. Ambient Intell. Humanized Comput., 10(5) (2019)1759–1768.

[15] Sofian Hamad and Taoufik Yeferny, Routing Approach for P2P Systems Over MANET Network, IJCSNS International Journal of Computer Science and Network Security, 20 (3) (2020).

[16] D. Helen; D. Arivazhagan, A stable routing algorithm for mobile ad hoc network using fuzzy logic system, International Journal of Advanced Intelligence Paradigms, 14(3/4) (2019) 248 - 259.

[17] Chalew Zeynu Sirmollo, Mekuanint Agegnehu Bitew, Mobility-Aware Routing Algorithm for Mobile Ad Hoc Networks, Wireless Communications and Mobile Computing, (2021).

[18] V. Ramesh, P. Subbaiah, and K. S. Supriya, Modified DSR (Preemptive) to reduce link breakage and routing overhead for MANET using Proactive Route Maintenance (PRM), Glob. J. Comput. Sci. Technol., 9(5) (2010) 124–129.

[19] T. D. Nguyen, J. Y. Khan, and D. T. Ngo, A distributed energy-harvestingaware routing algorithm for heterogeneous IoT networks, IEEE Trans. Green Commun. Netw., 2 (4) (2018) 1115–1127.

[20] L. Srinivasan, D. Selvaraj, T. P. Anish, "IoT-Based Solution for Paraplegic Sufferer to Send Signals to Physician via Internet," SSRG International Journal of Electrical and Electronics Engineering, vol. 10, no. 1, pp. 41-52, 2023.

[21] D. Selvaraj, T. P. Anish, "Outsourced Analysis of Encrypted Graphs in the Cloud with Privacy Protection," SSRG International Journal of Electrical and Electronics Engineering, vol. 10, no. 1, pp. 53-62, 2023.

[22] C. Chen and Y. Y. Cui, New method of energy efficient subcarrier allocation based on evolutionary game theory, Mobile Netw. Appl., 9 (2018)1–15, doi: 10.1007/s11036-018-1123-y.

[23] P.M. Joe Prathap, "Ensuring privacy of data and mined results of data possessor in collaborative ARM," Pervasive Computing and Social Networking. Lecture Notes in Networks and Systems, Springer, Singapore, vol. 317, pp. 431 – 444, 2022.

[24] Montoya, C. GuØret, J. E. Mendoza, and J. G. Villegas, ''A multi-space sampling heuristic for the green vehicle routing problem,'' Transp. Res. C, Emerg. Technol., 70 (2016) 113–128.

[25] Anish, T.P., Shanmuganathan, C., Dhinakaran, D., Vinoth Kumar, V., "Hybrid Feature Extraction for Analysis of Network System Security—IDS," Lecture Notes in Electrical Engineering, vol 1073. Springer, Singapore, 2023.

[26] S.M., Ananya, J., Roshnee, S.A. "IOT-Based Whip-Smart Trash Bin Using LoRa WAN," In: Jeena Jacob, I., Kolandapalayam Shanmugam, S., Izonin, I. (eds) Expert Clouds and Applications. ICOECA 2022. Lecture Notes in Networks and Systems, vol 673, pp. 277-288, 2023.

[27] V. V. Mandhare, V. R. Thool, and R. R. Manthalkar, QoS routing enhancement using metaheuristic approach in mobile ad-hoc network, Comput. Netw., 110 (2016) 180–191.

[28] J. Yang, M. Ding, G. Mao, Z. Lin, D.-G. Zhang, and T. H. Luan, Optimal base station antenna downtilt in downlink cellular networks, IEEE Trans. Wireless Commun., 18(3) (2019) 1779–1791.

___

[29] B. G. Sai Aswin, S. Vishnubala, D. Dhinakaran, N. J. Kumar, S. M. Udhaya Sankar and A. M. Mohamed Al Faisal, "A Research on Metaverse and its Application," 2023 World Conference on Communication & Computing (WCONF), RAIPUR, India, pp. 1-6, 2023.

[30] Dhinakaran, D., Selvaraj, D., Udhaya Sankar, S.M., Pavithra, S., Boomika, R., "Assistive System for the Blind with Voice Output Based on Optical Character Recognition," Lecture Notes in Networks and Systems, vol 492. Springer, Singapore, 2023.

[31] Taha, R. Alsaqour, M. Uddin, M. Abdelhaq, and T. Saba, Energy efficient multipath routing protocol for mobile ad-hoc network using the fitness function, IEEE Access, 5 (2017) 10369–10381.

[32] B. C. Latha, A. E. J. Anns and V. K. Sri, "Dam Management and Disaster Monitoring System using IoT," 2023 International Conference on Sustainable Computing and Data Communication Systems (ICSCDS), Erode, India, pp. 1197-1201, 2023.

[33] T. Kavya, S. Priyanka and P. P. Oviya, "A Way for Smart Home Technology for Disabled and Elderly People," 2023 International Conference on Innovative Data Communication Technologies and Application (ICIDCA), Uttarakhand, India, pp. 369-373, 2023.

[34] Monisha Ramakrishna Prabha; Thenmozhi Renganathan; Premkumar Mohan; Selvaraj Damodaran; Elumalai Govindarajan; Dhinakaran Damodaran, "Need for 6G: A survey," AIPConference Proceedings 2790, 020072, 2023.

[35] M. Keerthana, M. Ananthi, R. Harish, S. M. Udhaya Sankar and M. S. Sree, "IoT Based Automated Irrigation System for Agricultural Activities," 2023 12th International Conference on Advanced Computing (ICoAC), Chennai, India, pp. 1-6, 2023.

[36] M. Harini, D. Prabhu, S. M. Udhaya Sankar, V. Pooja and P. Kokila Sruthi, "Levarging Blockchain for Transparency in Agriculture Supply Chain Management Using IoT and Machine Learning," 2023 World Conference on Communication & Computing (WCONF), RAIPUR, India, pp. 1-6, 2023.

[37] K. Y. Kumar, N. J. Kumar, S. M. Udhaya Sankar, U. J. Kumar and V. Yuvaraj, "Optimized Retrieval of Data from Cloud using Hybridization of Bellstra Algorithm," 2023 World Conference on Communication & Computing (WCONF), RAIPUR, India, pp. 1-6, 2023.

[38] H.-C. Liu, J.-X. You, Z. Li, and G. Tian, Fuzzy Petri nets for knowledge representation and reasoning: A literature review, Eng. Appl. Artif. Intell., 60 (2017) 45–56.

[39] G. K. Monica, K. Haritha, K. Kohila and U. Priyadharshini, "MEMS based Sensor Robot for Immobilized Persons," 2023 International Conference on Innovative Data Communication Technologies and Application (ICIDCA), Uttarakhand, India, pp. 924-929, 2023.

[40] Thota, D. S. ., Sangeetha, D. M., & Raj , R. . (2022). Breast Cancer Detection by Feature Extraction and Classification Using Deep Learning Architectures. Research Journal of Computer Systems and Engineering, 3(1), 90–94. Retrieved from https://technicaljournals.org/RJCSE/index.php/journal/article/view/48

[41] S. K. Das and S. Tripathi, Intelligent energy-aware efficient routing for MANET, Wireless Netw., 24(4) (2018) 1139–1159.

[42] N. J. Kumar, S. S. Kamalesh and R. Abenesh, "Machine Learning System For Indolence Perception," 2023 International Conference on Innovative Data Communication Technologies and Application (ICIDCA), Uttarakhand, India, pp. 55-60, 2023.

[43] Joe Prathap P. M, Selvaraj D, Arul Kumar D and Murugeshwari B, "Mining Privacy-Preserving Association Rules based on Parallel Processing in Cloud Computing," International Journal of Engineering Trends and Technology, vol. 70, no. 3, pp. 284-294, 2022.

[44] S. K. Das and S. Tripathi, Energy efficient routing formation algorithm for hybrid ad-hoc network: A geometric programming approach, Peer-to-Peer Netw. Appl., (2018) 1–27, doi: 10.1007/s12083-018-0643-3.

[45] K. Sudharson, B. Alekhya, G. Abinaya, C. Rohini, S. Arthi and D. Dhinakaran, "Efficient Soil Condition Monitoring with IoT Enabled Intelligent Farming Solution," 2023 IEEE International Students' Conference on Electrical, Electronics and Computer Science (SCEECS), Bhopal, India, pp. 1-6, 2023.

[46] C. Cathrin Deboral, M. Ramakrishnan, "Safe Routing Approach by Identifying and Subsequently Eliminating the Attacks in MANET," International Journal of Engineering Trends and Technology, vol. 70, no. 11, pp. 219-231, 2022. https://doi.org/10.14445/22315381/IJETT-V70I11P224.